\documentclass[traditabstract]{aa} 
\usepackage{graphicx}
\usepackage{txfonts}
\newcommand{\zedited}{}
\begin{document}
   \title{The central density of R136 in 30~Doradus}


   \author{Fernando J. Selman
          \inst{}
          \and
          Jorge Melnick\inst{}
          }

   \institute{European Southern Observatory, 
              Alonso de Cordova 3107, Santiago\\
              \email{fselman@eso.org}
             }

   \date{Received 18 September 2012; accepted 13 February 2013}

 
  \abstract
   {
	The central density, $\rho_0$, of a stellar cluster is an important physical
	parameter for determining its evolutionary and dynamical state.
	How much mass segregation there is or whether the cluster has undergone core collapse both
	depend on $\rho_0$. We reanalyze the results of a previous paper that gives
	the mass density profile of R136 and combine them with both a conservative upper
	limit for the core parameter and a more uncertain recent measurement.
    	We thus place a lower limit on $\rho_0$~under reasonable and defensible assumptions
	about the IMF, finding $\rho_0 \geq 1.5\times10^4 M_\odot/pc^3$\ 
	for the conservative assumption $a < 0.4 {\rm pc}$ for the cluster core parameter.
   	If we use the lower, but more uncertain value $a = 0.025 {\rm pc}$,
	the central density estimate becomes greater than $10^7 M_\odot/pc^3$.
	A mechanism based on the destruction of a large number of circumstellar disks
	is posited to explain the hitherto unexplained increase in reddening
	presented in that same work.
   }


   \maketitle
%

\section{Introduction}

Owing to its nearness and extreme nature, Radcliffe~136 (Feast, Thackeray, and Wesselink \cite{feast1960})
has been not only a template for other more extreme and distant
starburst clusters, but also a good example of how a typical
galactic globular cluster might have looked soon after its birth. To obtain
its physical characteristics in a way firmly rooted in observations
has been an important goal as the many studies of this interesting
object attest (
Melnick \cite{melnick1985};
Weigelt and Baier \cite{weigelt1985};
Campbell et al. \cite{campbell1992};
Parker and Garmany \cite{parker};
Malumuth and Heap \cite{malumuth};
Hunter et al. \cite{hunter};
Brandl et al. \cite{brandl};
Hunter et al. \cite{hunterb};
Walborn and Blades \cite{walborn1997};
Massey and Hunter \cite{massey1998};
Andersen et al. \cite{andersen2009};
Bosch et al. \cite{bosch2009};
Campbell et al. \cite{campbell2010};
Crowther et al. \cite{crowther2010};
de Marchi et al. \cite{demarchi2011};
H\'enault-Brunet et al. \cite{henault2012}, and the whole series of VLT-FLAMES Tarantula Survey papers;
Sabbi et al. \cite{sabbi}).
Selman et al. (\cite{selman1999b}, henceforth SMBT) studied
the IMF of R136 and provided several of its physically relevant
parameters, including the mass density profile of the cluster.
Three very important results of that paper gave insight
into the age structure of the cluster, the normal nature
of its IMF, and the scale-free character of the mass density
profile between 0.4 and 10 pc.

The cuspy profile and the small core radius has been used to posit a post-core-collapse
(PCC) state for this cluster (Campbell et al.\cite{campbell1992};
but see opposite views by Malumuth and Heap \cite{malumuth},
Brandl et al. \cite{brandl}, and Mackey and Gilmore \cite{mackey2003}).
PCC clusters are characterized by a high central density
and a density profile that can be modeled by a normal King
profile, which shows a break and turns into a power law near
the center. High central density means, in this context, that
the relaxation time is shorter than the age of the system.
Observations of Galactic globular clusters show that they
can be separated into ``core-collapsed'' and non-core-collapsed
depending on whether their photometric profile turns into
a power law near the center (Harris \cite{harris}; McLaughlin and van
der Marel  \cite{mclaughlin2005}; Chatterjee et al. \cite{chatterjee}
and references therein).
The cluster R136 has been claimed to have the characteristics
of a PCC cluster: a cuspy density profile
(Mackey and Gilmore \cite{mackey2003}), and a large number of runaway stars
(see Fujii et al. (\cite{fujii}) and references therein). Nevertheless,
the time scale for core collapse is too long if we believe
the value of $3\times10^4 M_\odot/{\rm pc}^3$~determined
by Mackey and Gillmore (\cite{mackey2003}) for the central density.
Fujii et al. (\cite{fujii}) invoke the fact that smaller
clusters evolve faster, thereby speeding up the evolution via
hierarchical merging of smaller substructures. This idea
has been backed up by the recent work of Sabbi et al.~(\cite{sabbi}).
Using theoretical isochrones for MS and PMS evolution, and HST optical
and NIR photometry, they find that the central part of the cluster
is very young with ages below 1My, while an overdensity $\sim$5.4~pc
to the northeast of the center is closer to 2.5~My.
This led them to propose that R136 is a double cluster that is
currently interacting. Similar ages were found in SMBT, but with
fits to the Geneva tracks in the upper part of the HR diagram.
That similar age structure is found by using these two sets of
tracks give credibility to both of these results. 

Such a complex age structure is quite a complication when
modeling to convert magnitudes and colors into masses.
If the average age of the stellar population depends
on radius, this will result in systematic errors in our
estimates of physical parameters unless variable ages
are allowed in the modeling. Most work on R136 so far
assume simple stellar populations. The only work that we are
aware of that does not is SMBT, which also considers
variable reddening determined in a star-by-star basis,
and does a full completeness analysis, which is fundamental
when working in the optical bands.

Using ground-based observations in combination with the
HST work of Hunter et al.~(\cite{hunter,hunterb}), SMBT
determined the density profile of the cluster showing that between 0.4 - 10~pc
it is represented well by a single power law,
giving explicit expressions. Although a lower limit to the central
density of the cluster can be derived from those numbers,
it was not given explicitly in the paper. In this work we
give the explicit results. If we use the same core parameter as used
by Mackey and Gilmore (\cite{mackey2003}) we find a striking similarity
between our result and that of those authors. Furthermore, using the
recent estimates of the core parameter by Campbell et al.~(\cite{campbell2010}),
we find that the central density is considerably higher than the previous estimate,
making core collapse a virtual certainty for this cluster, or an indication
that the latter estimate of the core parameter is wrong.


\section{Radial profile, central  density, and total mass in R136}

SMBT give the following relation for the stellar number density
normalized to 1~pc, derived counting stars with masses between
$10M_\odot < M < 40M_\odot$:

\begin{eqnarray}
\rho_n(r) & = & 9.8\ {\rm stars\over pc^3}\ \left( {1{\rm pc} \over r}\right)^{2.85},
\label{eq:pap1}
\end{eqnarray}where $r$\ is the distance to the cluster center in parsecs.
As previously noted,
a strong effort was made by SMBT to correct for incompleteness, and
although the work was based on optical data, Equation~(\ref{eq:pap1}) should
not be affected by differential reddening and should be complete in the
specified mass range.

A form often used to parametrize the stellar density of a star cluster
was given by Elson, Fall, \& Freeman (\cite{elson}):
\begin{eqnarray}
\rho(r) & = & \rho_0 \times {1 \over \left(1+ \left({r\over a}\right)^2\right)^{{\gamma+1\over 2}}},
\label{eq:eff}
\end{eqnarray}where $a$ is the core parameter\footnote{The core parameter in the Elson,
Fall and Freeman form is related to the King profile core radius by $r_c = a\sqrt{2^{2/\gamma}-1}.$
This formula is incorrectly quoted in Campbell et al. (\cite{campbell2010}),
but correctly given in Mackey and Gilmore (\cite{mackey2003}). We also note that $r_c$~is
equivalent to the $r_0$~in the King (\cite{king1962}) model only when it is much smaller
than the tidal radius. This is not valid for low-concentration clusters, but certainly
valid for R136.},
and $\rho_0$~is the central density. This form is suitable for clusters that are
young enough not to exhibit tidal truncation. We note that the core parameter \emph{a} is usually
obtained by fits to the 2-D profile, but if the fit is a good one, then the value
thus obtained is the same as for the cluster 3-D density profile.
Given a core parameter, then and only then
can we calculate $\rho_0$. If we only have valid measurements in the power-law region,
where $r >> a$, we can only obtain a lower limit to $\rho_0$.
Thus, the challenge in determining the central
density is really a challenge in determining the
core parameter. There are basically two methods of
determining this parameter: fits to star counts, and fits
to the integrated light profile. Brandl et al. (\cite{brandl}) cautions
that determining the core parameter using surface brightness
profiles is not reliable, since the light can be dominated by a few very
bright stars (as is indeed the case near the center of R136), and mimic a power-law
cusp. A similar caveat can be found in Mackey and Gilmore (\cite{mackey2003}).

\begin{figure}
   \centering
   \includegraphics[angle=-90,width=9cm]{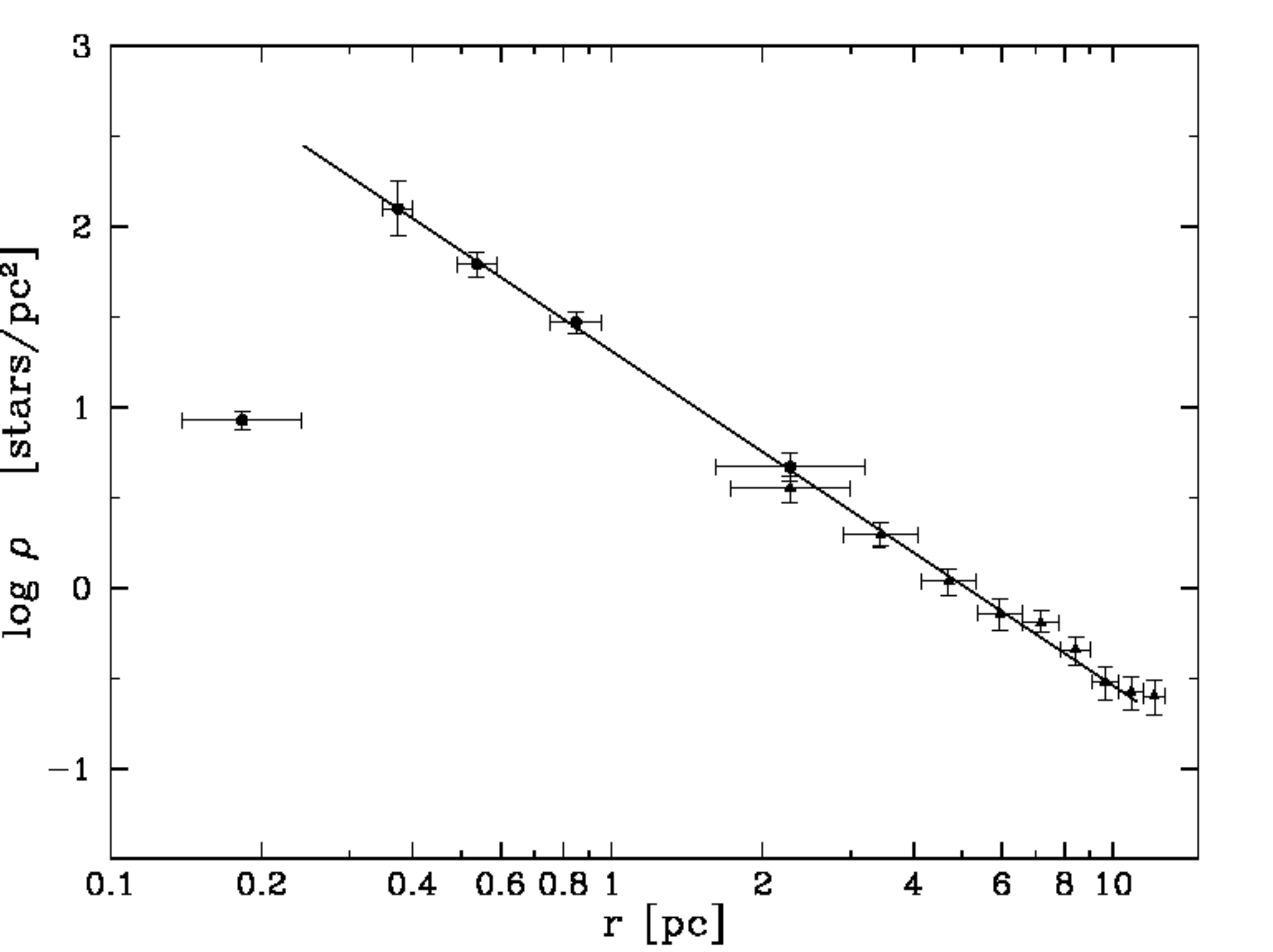}
   \caption{ Radial density profile for the stars with $10M_\odot < M < 40M_\odot$.
		For the innermost bins we have calculated the points
		using the Hunter et al. (\cite{hunter}) data following
		the procedure described in {\zedited SMBT}. The line is
		a power law with exponent -1.85. The point at the smallest
		radius is strongly affected by incompleteness. Reproduced
		with permission from SMBT.
               }
              \label{SMBT_Fig13}%
    \end{figure}

We can  write the results for the stellar volume density
found in SMBT using the Elson, Fall, \& Freeman (\cite{elson}) form  for an arbitrary
value of $a$\ but fitting the SMBT power-law data:
\begin{eqnarray}
\rho_n(r) & = & 9.8 {\rm stars\over pc^3}\times \left({1 pc\over a}\right)^{1.85+1}\ {1 \over \left(1+ \left({r\over a}\right)^2\right)^{{1.85+1\over 2}}},
\label{eq:SMBT_EFF}
\end{eqnarray}showing a clear power law of slope $\gamma = 1.85$, without signs of a core between 0.4 - 10 pc.
Figure~\ref{SMBT_Fig13} reproduces Figure~13 from
SMBT, which gives a profile totally determined counting
stars in \emph{mass} bins. The innermost point suffers from
strong incompleteness.

Using the value $a=0.4~{\rm pc}$\ as an upper limit we can write the previous result as
\begin{eqnarray}
\rho_n(r) & \geq & 133 {\rm stars\over pc^3} {1 \over \left(1+ \left({r\over 0.4~{\rm pc}}\right)^2\right)^{{1.85+1\over 2}}}.
\label{eq:SMBT_EFF_a0p4}
\end{eqnarray}This expression is valid for stars with M in the range
$10M_\odot < M < 40M_\odot$.  To convert this number density profile
to a mass profile, we need the spectrum of stellar masses to extrapolate
the density from the given mass range to the full mass range.
We choose to use the spectrum of masses at birth, i.e. the IMF,
instead of the present-day mass function, although it does not
make much of a difference given the young age of the central
parts of R136. At 1.2~My the most massive stars have lost less
than 4\% of their mass, {\zedited and at 2~My this figure is still only 20\%.}
The lowest mass for which an IMF has been given for R136 so far is
1$M_\odot$ (Andersen et al. \cite{andersen2009}).
These authors found consistency with a Salpeter slope down to this mass limit.
We used this result as a check that the IMF does not show
a break down to this mass limit but used the SMBT IMF slope extrapolated
down to this mass, because it was determined in a rigorous and
self-consistent way.  We thus use a Kroupa IMF (Kroupa \cite{kroupa}) modified in
the high mass range to have the  slope determined in SMBT ($\Gamma=1.17$),

\begin{eqnarray}
\xi(m) & = & \left\{ \begin{array}{c}
                \left({m\over 0.08}\right)^{-0.3} \hfill 0.01 <= m < 0.08 \\
                \left({m\over 0.08}\right)^{-1.3} \hfill 0.08 <= m < 0.5 \\
                \left({0.5\over 0.08}\right)^{-1.3}\times\left({m\over 0.5}\right)^{-2.17} \hfill 0.5 <= m < 120.
                \end{array} \right.
\label{eq:IMF}
\end{eqnarray}
With this IMF the fraction of stars with masses between $10M_\odot < M < 40M_\odot$
is 0.3967\%. The fraction in the mass range $1M_\odot < M < 120M_\odot$\ is 7.286\%,
and the average mass is 3.85M$_\odot$.
Thus, if we choose to integrate only down to 1$M_\odot$, we
estimate the central mass density limit,

\begin{eqnarray}
\rho_0 & \geq & 9.4\times10^3 {M_\odot\over {\rm pc}^3}.
\label{eq:SMBT_rhonot_1msun}
\end{eqnarray}

The result in Equation~\ref{eq:SMBT_rhonot_1msun} has
been derived almost directly from observations. There
are modeling uncertainties, and an extrapolation
from 10$M_\odot$~down to 1$M_\odot$, but the latter supported by observations.
To extend the result to even lower masses where the mass function
has not been determined observationally is somewhat risky.
Depending on the actual dynamical state of
the cluster, we could have the effects of
mass segregation invalidating the result.
The mass density profile we obtain with this extrapolation of the IMF
to the full $0.01M_\odot < M < 120M_\odot$~range where the average mass
is 0.46$M_\odot$~is given by
\begin{eqnarray}
\rho(r) & \approx & 1.5\times10^4 {M_\odot\over\rm pc^3} {1 \over \left(1+ \left({r\over 0.4~{\rm pc}}\right)^2\right)^{{1.85+1\over 2}}}.
\end{eqnarray}\noindent Mackey and Gilmore (\cite{mackey2003}) determine an
upper limit to the core parameter of $a<0.32~pc$, and with this value we
can estimate a lower limit to the central density of $2.8\times10^4 {M_\odot/pc^3}$,
in excellent agreement with the value of $3\times10^4 {M_\odot/pc^3}$~given by
those authors. Because of the IMF extrapolation we have not
preserved the inequality sign in this equation, which was quite certain
for Equation~\ref{eq:SMBT_rhonot_1msun}.

These numbers can also be used to determine a total mass. The complication
here is that the slope of  the density profile results in an ever increasing
total mass as one goes to larger radii. Here we choose to cut the integration
at a radius of 10~pc, the distance at which the star counts become
noisier due to the presence of a number of substructures.
The same slope value that diverges for large radii ensures convergence
when integrating from the center. Thus, we will give our estimates
of the mass of the cluster as a range where the lower total mass limit
is given by the upper limit of the core parameter, and the upper total mass
limit is given for a power law all the way to the center. With these
assumptions we can then write
\begin{eqnarray}
M_{tot} = 4\pi a^3\rho_0 \int_0^{R/a} {u^2du\over (1+u^2)^{\gamma+1\over 2}} = 4\pi a^3\rho_0 F\left({R\over a}\right) ,
\end{eqnarray}where we have defined
\begin{eqnarray}
F\left({R\over a}\right) = \int_0^{R/a} {u^2du\over (1+u^2)^{\gamma+1\over 2}},
\end{eqnarray}and in what follows we integrate numerically. We
note that for $\gamma > 2$ the integral converges in
the limit $R/a\rightarrow\infty$. For $\gamma=2$ it diverges
logarithmically.

With these definitions the constraint to the total mass becomes
\begin{eqnarray}
4\pi a^3\rho_0 F\left({R\over a}\right) < M_{tot} < 4\pi a^3\rho_0 F\left({R\over a}\right) {{{1\over 2 - \gamma}({R\over a})^{2-\gamma}}\over F\left({R\over a}\right)}
\end{eqnarray}and for R136 using $a<0.4~{\rm pc}$,
\begin{eqnarray}
4.6\times10^4 M_\odot < M_{tot} < 1.3\times10^5 M_\odot,
\end{eqnarray}or in terms of total number of stars $N_{tot}$,
\begin{eqnarray}
10^5 < N_{tot} < 2.8\times10^5.
\end{eqnarray}

\section{Discussion}

As noted in the previous section, the estimate of the central
mass density depends very strongly on the limits  put
on the core parameter. Because it was determined by star counts,
we feel that the limit given in SMBT is a very conservative one.
{\zedited Other authors} have placed a much more stringent constraint by
analyzing the light profile (e.g. Campbell et al.~\cite{campbell1992};
Andersen et al.~\cite{andersen2009}). One of the latest such studies,
and one that reaches similar conclusions to other studies that use
the surface brightness method, is that of Campbell et al. (\cite{campbell2010}).
They used the multiconjugate adaptive optics instrument (MAD) at
ESO's Melipal 8-m telescope to perform star counts and surface photometry
in H and K, on frames with  typical Strehl ratios in K of 15 - 25\%.
Although there is little leverage for disentangling
masses and ages in the infrared bands, they are less affected by reddening so with proper
care a good estimate of the radial profile can be obtained in the
case of a \emph{simple stellar population.} These bands
are also separate from the peak emission for very massive stars so the cluster's
light profile is less affected by the presence of a few very massive stars.
The magnitude limit of the MAD data corresponds to approximately 5$M_\odot$.
Within approximately $r<2 pc$\ they use
the light profile integrated in annuli, while for $ r > 0.7 pc$\ they
use star counts. They merge the two profiles into a single one using the
area of overlap to normalize the two sets. They find $\gamma=-1.6\pm0.1$,
and $a = 0.025 pc$. Because of the presence of radially dependent
stellar populations with multiple ages, we only use
their value for the core radius, while for the slope we use the
value determined in SMBT where allowance has been made for multiple epochs of
star formation.

The issues complicating the determination of the core
parameter using integrated light profiles are not addressed
at all by Campbell et al. (\cite{campbell2010}),
but are fundamental in these kinds of studies.
Their Figure~20 shows that the power-law slope is virtually
the same for the segment derived with star counts and the segment derived
by the surface brightness fit. This might indicate that the surface brightness
is a good proxy for mass surface density. Nevertheless, we must
point out that the very bright central sources could give a similar
signal due to scattering of light in the optical elements of the instrument.
It has been known for a long time that the profile of a stellar image contains
a kernel and a power-law part of index $\sim-2$ (de Vaucoulerus \cite{devaucouleurs1958}; King \cite{king1971}).
The origin of this power-law halo has been ascribed to optics
imperfections and dust. According to Kormendy~(\cite{kormendy1973}) the slope
is flatter at $\sim-1.54$. This index is close to the -1.6 index
determined with MAD so scattered light might mimic the cluster
light profile. This shallow-slope power law is expected to dominate
a few seeing radii away from the core of the PSF. Nevertheless, between the core
of the PSF and this radius, we expect a much steeper
uncompensated halo that falls off with a Moffat $\beta\sim 11/3$,
approximately as $r^{-3.7}$ (Moffat~\cite{moffat1969}; Roddier~\cite{roddier1995};
Racine~\cite{racine1999}).

The results of Campbell et al. of $a = 0.025~pc$,
imply moderately larger bounds for the total mass
and number of stars given by 
\begin{eqnarray}
\begin{array}{rcccl}
7.5\times10^4 M_\odot & < & M_{tot} & < & 1.3\times10^5 M_\odot, \\
1.6\times10^5         & < & N_{tot} & < & 2.8\times10^5,
\end{array}
\label{eq:Ntot_dense}
\end{eqnarray} and central density of
\begin{eqnarray}
\rho_0 & \geq & 4.0\times10^7 {M_\odot\over {\rm pc}^3},
\label{eq:rhonot_dense}
\end{eqnarray}
which is much higher than any other estimates of
the R136 central density.

Is R136 a PCC cluster? If we take the MAD data at face
value, the cluster must be in a PCC state since the relaxation time for
a cluster with such a central density is only $\sim8\times10^3$y,
and the time to core collapse is approximately 15 times
this value (using Binney and Tremaine (\cite{binney}),
as presented in Equation~8 by Mackey and Gilmore (\cite{mackey2003})),
or $\sim1\times10^5$y, considerably below the current estimate of $10^6$y
for the age of the cluster.

Two other subtleties related to relaxation
and core collapse must be pointed out. First, the identification
of ``core-collapsed'' clusters in the galaxy rests in identifying
a \emph{break} in the inner part of the light profile. 
When properly studied, that is by star counts, R136 shows a scale-free
distribution from 10~pc all the way down to at least 0.4~pc. Second,
the relaxation times used in the literature assume a single
mass component, while for this very young cluster we have a
spectrum of masses spanning at least three orders of magnitude.
In such a case the relaxation time is diminished by the ratio
of the lowest mass to the highest mass star. We do not delve
deeper in this subject because the physics of core-collapse is beyond
the scope of this work.

   \begin{figure}
   \centering
   \includegraphics[angle=-90,width=9cm]{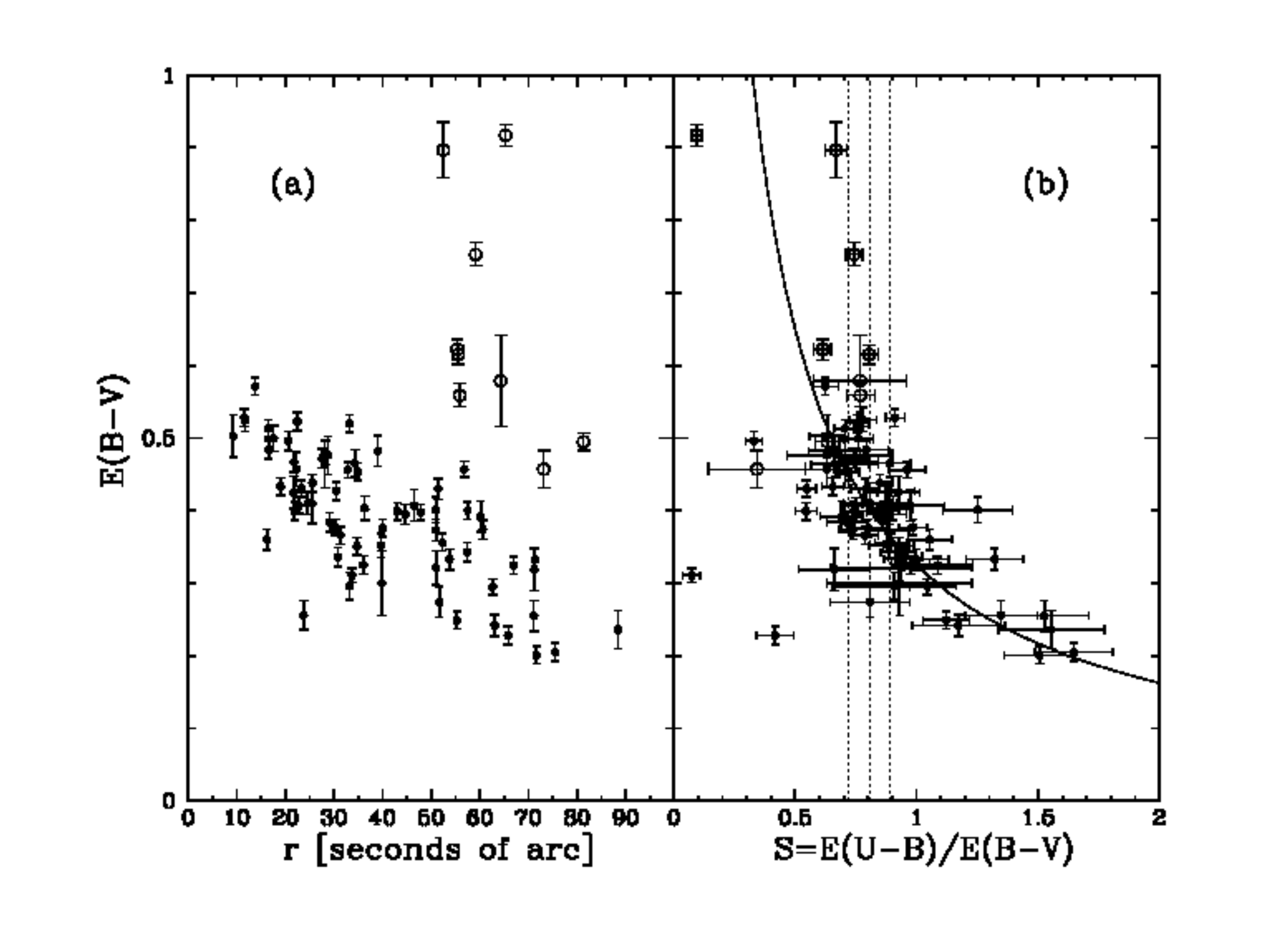}
   \caption{(a) Radial profile of color excess, as determined
		in Selman et al~(\cite{selman1999}). (b) Colour
		excess versus selective extinction parameter S. 
		Reproduced with permission from Selman et al. (\cite{selman1999}).
               }
              \label{fig:redSEBMV}%
    \end{figure}

A central density  as high as implied by
Equation~\ref{eq:rhonot_dense} would have other consequences.
Selman et. al. (\cite{selman1999b}), the first paper of the series
and dedicated in part to the study of the reddening distribution
in R136, {\zedited found a puzzling trend in reddening increasing toward
the center of the cluster.} Their Figure~6(a),
reproduced here as Figure~\ref{fig:redSEBMV} shows clearly that
trend. The figure shows that in the inner regions most of the stars
show a somewhat limited amount of variable extinction that increases
toward the center. In the outer parts {\zedited more than half} of the stars
follow the same trend, while another group, probably associated
with the nebulosity that surrounds the cluster, show a much higher
variation of extinction.  The amount of dust associated with the
inner component was estimated at 30-60$M_\odot$~within 15~pc,
and its origin was a mystery.

The details of the calculation were not given in SMBT, so we give them
in Appendix~\ref{appendix}. From Figure~\ref{fig:redSEBMV} we see
that the extinction increases from values near 0.2-0.3 at 80\arcsec
from the center reaching values slightly in excess of 0.5 near the center.
We assume that the former values sum the contributions from the Milky Way,
the LMC and the larger 30 Doradus environment, and we assume the values
above 0.2-0.3 to come from a component associated to R136 itself.
Thus, using $E_{B-V}\approx 0.1 - 0.2$~for the extinction associated
with R136, we find $M_{\rm dust}\sim 40 - 80 M_\odot$~within 15~pc
from the cluster center, or $M_{\rm dust}\sim 18 - 36 M_\odot$~within
10~pc from the cluster center.

In SMBT, we stated our surprise to see dust in such an
extreme environment. Any dust particle near the center of the
cluster would, if not destroyed, be removed very rapidly. Thus,
a source of dust is needed to replenish it. A natural source
of matter is the mass loss from the massive stars.
It is only the stars with masses above 15M$_\odot$ that lose
significant amounts of mass, and the total amount of mass loss
depends strongly on the age of the system as the rate
increases rapidly with age. According to SMBT the central
cluster stars belong to the first burst with ages below 1.5~My.
A more recent estimate of the age of the central stars of the
cluster is given by Crowther et al.~(\cite{crowther2010}). These
authors used rotating models and UV and IR spectroscopic observations
of the central WN stars to constraint the age to 1.7$\pm$0.2~My.
We use for our estimate an upper limit of 2~My with the
evolutionary models without rotation of Meynet et al.~(\cite{meynet1994}).
These authors doubled the mass loss rates of de Jaeger et al. (\cite{dejager1988}).
With this artificial increase in the mass loss rate, these authors were
able to explain the number ratio of WR stars to O stars, which otherwise
would have come too low. With the current state of the art rotating
models they are able to produce WR stars even if there is no wind!
The new models use the mass loss rates for O stars by Vink et al.~(\cite{vink2001}),
and for WR stars by Nugis and Lamers~(\cite{nugis2000}) that have
mostly revised the rates downward (Maeder and Meynet~\cite{maeder2012}).
Thus, using the Meynet et al.~(\cite{meynet1994}) 2~My isochrone
for Z=0.008, weighted by the IMF in Equation~\ref{eq:IMF}
we can calculate an upper limit to the total mass
loss due to stellar evolution that is 0.35\%.
If we set the total mass of the cluster to 10$^5M_\odot$,
then the total mass loss corresponds to 350~M$_\odot$.
Although a natural source of matter is the mass loss from
the massive stars, it is not clear that they are a natural
source of dust. To convert this to an amount of dust, we
need the uncertain gas-to-dust ratio in the wind of such hot stars.
The amount of gas from mass loss, if it were to have the gas-to-dust ratio
of 347 found by Pei~(\cite{pei1992}) for the LMC would correspond to only 1.0~M$_\odot$,
more than an order of magnitude below what we see.
Using 100 for the gas-to-dust ratio results in only 3.5~M$_\odot$
of dust. Thus, in SMBT we realized that
we needed a different source of dust. Dust around
massive stars, B-types, and WR had been found in the past
(Geisel~\cite{geisel1970}, Allen and Swings~\cite{allen1972}),
which was later ascribed to colliding winds (Usov~\cite{usov1991}),
so we speculated that this dust could be produced in colliding winds
of massive stars. Some work has been done in colliding winds since that time,
and it has become clear that to produce dust we need
a binary with a WC star as a member (Crowther \cite{crowther2003}).
There is only one WC star near the center of R136 in projection,
Brey 83 (Breysacher \cite{breysacher1981}).
This star is  found at approximately 3~pc from the cluster center, and given the
age structure of the cluster, it is probably farther away from the center.
Even so, one star is not enough to explain the large amount of dust found.
We would like to point out another possibility here.

If the IMF is normal, which it appears to be down to the lowest
masses studied, we can assume that the star formation mechanism must also
be normal, and we can expect the presence of circumstellar disks.
The observational results are somewhat difficult to interpret.
Rubio et al.~(\cite{rubio1998}) used JHKs photometry to detect
infrared excess sources around the 30 Doradus nebula. Most of
the sources with excess are more than 5~pc away from the central
cluster. These authors interpret this emission as coming
from YSOs whose formation was triggered by the central cluster.
Maercker and Burton~(\cite{maercker2005}) use L-band observations
{\zedited with the 60-cm South Pole InfraRed Explorer (SPIREX) Telescope}
to detect L-band excess sources that they
interpret as possessing circumstellar disks. They find a
disk fraction of 42\%. These observations are limited to
the brightest sources, and the resolution does not permit
studying the central regions of the cluster. It is likely
that these sources are very bright YSOs that are still embedded
in their dust shells. Stolte et al.~(\cite{stolte2010}) studied
the galactic star-burst Arches cluster near the center of
the galaxy using L-band excess to find candidate
disk sources. In addition they confirm the disk nature of these
excess sources using SINFONI VLT observations to detect
the signature of rotation in the CO bandhead emission
(Bik and Thi~\cite{bik2004}). They find that the disk fraction
in Arches increases from 2.7\%$\pm$1.8\% in the cluster
core (within 0.16~pc from the center), to 5.4\%$\pm$2.6\%
for $0.16pc<r<0.3pc$, reaching 9.7\%$\pm$3.7\% outside.
The overall disk fraction for the B stars in the Arches
cluster is 6\%, much lower than what would expect
from the disk fraction age relation (Haisch et al~\cite{haisch2001}).
They explain the decrease in disk fraction toward the
center by either UV radiation destruction, winds, or
the tidal destruction mechanism of Olczak et al. (\cite{olczak2012})
that has the important characteristic, in the context of
this work, of destroying the disk without destroying the dust.

In a recent series of papers, Olczak et al. (\cite{olczak2012}, and references
therein) have proposed, based upon numerical simulations,
that in an environment as dense as the core of the Arches cluster,
where the density is expected to be $>10^5M_\odot/pc^3$,
encounters can destroy up to one third of the circumstellar disks.
Olczak, Pfalzner, and Eckart~(\cite{olczak2010}) posit
a number density of 10$^5pc^{-3}$\ as the threshold above
which disk destruction occurs rapidly.
If Equation~\ref{eq:rhonot_dense} represents the physical state
of the cluster, then the disk destruction process might
be responsible, at least in part, for the inner extinction that we
have measured. According to recent studies, disks are formed
in all environments, varying only the lifetimes of the disks,
which are much shorter for massive stars where they are destroyed
by photo-evaporation (See Williams and Cieza \cite{williams2011}
and references therein). There is a loose linear relation
between disk mass and stellar mass. The \emph{median} disk mass
is approximately 1\% of the parent star mass, and the variation
is approximately $\pm$0.5~dex. For our estimate we need the
arithmetic mean, which for stellar masses between 0.015-4~M$_\odot$
has been found to vary between 2\% to 5\% of the parent stellar
mass for different galactic star forming regions studied
(Andrews et al. \cite{andrews2010}; Scholz et al.~\cite{scholz2006},
Manning and Sargent~\cite{manning2000}). Thus, if we assume a total
mass of 10$^5M_\odot$ for the cluster, we expect the total mass
in disks to be 2000-5000~M$_\odot$.
Using a gas-to-dust ratio of 100 implies 20-50~M$_\odot$~of dust,
while using the gas-to-dust ratio of the 30 Doradus region of
the LMC results in 6-15~M$_\odot$.
Thus, to explain the inner reddening of R136 with dust
coming from the destruction of circumstellar disks requires
(1) that most of these disks were destroyed
recently, within a few 10$^5$y or else the gas would have been
removed from the central region; and (2) the gas-to-dust ratio
must be closer to 100 than to the value of 347 measured elsewhere
in the 30 Doradus region. If both of these conditions are not
met, then an alternative origin must be sought.

\section{Conclusions}

In this work we estimate a conservative limit to
the central mass density of R136 of $1.5\times10^4 M_\odot/{\rm pc}^3$
for $a<0.4 {\rm pc}$. From this we estimate that the total
mass of the cluster enclosed within 10~pc must be in the
range
\begin{eqnarray*}
4.6\times10^4 M_\odot < M_{tot} < 1.3\times10^5 M_\odot,
\end{eqnarray*}or in terms of total number of stars $N_{tot}$,
\begin{eqnarray*}
10^5 < N_{tot} < 2.8\times10^5.
\end{eqnarray*}In this case the observed scale-free profile must
be in place at the moment of formation because relaxation and core
collapse have not had time to act in this young object.

On the other hand, if we use the recent estimate of the core
parameter by Campbell et al. (\cite{campbell2010}), we get
the moderately larger bound for the total mass
\begin{eqnarray*}
7.5\times10^4 M_\odot < M_{tot} < 1.3\times10^5 M_\odot,
\end{eqnarray*}
and the more extreme constraint for the central density
\begin{eqnarray*}
\rho_0 & \geq & 4.0\times10^7,
\end{eqnarray*}which would imply that core collapse becomes a certainty.

Such high central density can have other effects. We have shown that
it is unlikely that the reddening distribution, observed by SMBT to increase toward
the inner parts of the cluster, could come from
normal mass loss from massive stars. It is posited in this work
that a sizeable fraction of this dust could come from the recent destruction
of most of the circumstellar disks in R136.

\begin{acknowledgements}
The authors would like to thank C. Olczak for the estimate
of the number of disks needed to be destroyed to explain
the total mass of dust implied in Selman et al. (\cite{selman1999}).
The speculation that the color excess seen in the inner parts
of R136 is due to such a process lies only with the authors.
C. Olczak also pointed out to us the effect that a spectrum
of masses can have on the relaxation time of a stellar system.
We would also like to thank the anonymous referee, whose constructive
criticism helped improve the paper. We would also like to
thank Dave Jones and the laguage editor, Joli Adams, for help with the English.
\end{acknowledgements}

\appendix
\section{The total mass of dust from the observed E(B-V)}
\label{appendix}
For our estimates we follow Purcell (\cite{purcell1969}), Spitzer (\cite{spitzer1978}),
and Pei (\cite{pei1992}). This last author is interested
in the gas-to-dust ratio, and all his derivations were done in terms
of that parameter. However, the extinction can be directly converted
into a mass column density of dust based solely on the assumed dust model.
Spitzer (\cite{spitzer1978}) gives an expression for the mass of dust
as a function of E(B-V), but valid only for the Milky Way and in terms
of the average extinction per kpc.  We therefore redo the more modern Pei (\cite{pei1992})
calculation to directly get the dust column density implied by the observed E(B-V).

The extinction optical depth at wavelength $\lambda$, $\tau(\lambda)$, is given by
\begin{eqnarray}
\tau(\lambda)=\sum_X\int_{a_{min}}^{a_{max}}da\pi a^2n_X(a)Q_X(\lambda,a)
\end{eqnarray}where the sum is over species X, graphite, and silicate; $a$~is
the size of a grain; $n_X(a)$~is the distribution of grain sizes
of type X; a$_{min}$~and a$_{max}$ are the
lowest and largest grain sizes; and $Q_X(\lambda,a)$~is the
extinction efficiency factor for grains of type X. The distribution of grain sizes
is that of Mathis et al. (\cite{mathis1977})\footnote{Note a slight departure from
the Pei (1992) notation. We use $n_X(a)$~for the distribution of grain
sizes instead of $N_X(a)$, {\zedited and use  $N_X$~only as the normalization constant.}}:
\begin{eqnarray}
n_X(a)da = \left\{ \begin{array}{ll}
         N_X (a/a_V)^{-\beta}d(a/a_V) & \mbox{if $a_{min} < a < a_{max}$};\\
         0 & \mbox{otherwise},\end{array} \right.
\end{eqnarray}with $\beta=3.5$, $a_{min}=0.005\mu m$,
and $a_{max}=0.25\mu m$. Then, a$_V$ is given by
\begin{eqnarray}
a_V^3\equiv N_X^{-1}\int_{a_{min}}^{a_{max}}da n_X(a)a^3,
\end{eqnarray}which for this grain size distribution results in $a_V=0.737\mu m$.
With these definitions they write
\begin{eqnarray}
\tau(\lambda)/\tau_B = \sum_X r_X \int_{a_{min}}^{a_{max}} d(a/a_V) (a/a_V)^{2-\beta}Q_X^{ext}(\lambda,a)
\end{eqnarray}where $Q_X^{ext}(\lambda,a)$~is the extinction efficiency
factor of grains of type X given by Draine and Lee (\cite{draine1984}).
They have defined r$_X\equiv\pi a_V^2 N_X/\tau_B.$ They use this expression,
together with a dust grain model,
to fit the extinction data to the Milky Way, the LMC, and the SMC to find the
value of r$_X$~that fits the data. With these definitions the column density
of dust, $\Sigma_d$, is given by
\begin{eqnarray}
\Sigma_d &  =  & \sum_{X=C,S} N_X{4\over 3}\pi a_V^3\rho_X\\
         &  =  & {4\over3}(r_C\rho_C + r_S\rho_S)a_V\tau_B\\
         &  =  & {4\over3}{a_V\over1.086}(r_C\rho_C + r_S\rho_S)(1+R_V)E(B-V)
\end{eqnarray}where $\rho_C=2.26\,g\,cm^{-3}$~and
$\rho_S=3.3\,g\,cm^{-3}$~are the mass density of the graphite and silicate grains, and
R$_V$ is the total-to-selective extinction parameter, which has the value 3.16 for the LMC.
For the LMC the extinction curve is best fit with $r_X$~values given by $r_C=0.018$~and
$r_S=0.083$. The values for the Milky Way and the SMC are somewhat
different and can be found in Pei (\cite{pei1992}), and can be used to finally obtain
\begin{eqnarray}
\Sigma_d = \left\{ \begin{array}{ll}
	    0.40\,E(B-V)\, {M_\odot/pc^2} & \mbox{for the Milky Way};\\
            0.57\,E(B-V)\, {M_\odot/pc^2} & \mbox{for the LMC};\\
	    0.67\,E(B-V)\, {M_\odot/pc^2} & \mbox{for the SMC}.\end{array} \right.
\end{eqnarray}Thus, an average value $E(B-V)\sim 0.1 - 0.2$~implies, in the LMC, a total
mass of 40-80~M$_\odot$~within 15~pc, and 18-36~M$_\odot$~within 10~pc.


\end{document}